# Direct and inverse scattering problems for spherical Electromagnetic waves in chiral media

Christodoulos Athanasiadis[1] and Nikolaos Berketis[2]

*Department of Mathematics, University of Athens, Panepistemiopolis, GR 15784 Athens, Greece*

## SUMMARY

We study the direct and inverse scattering problems when the incident electromagnetic field is a time harmonic point- generated wave in a chiral medium and the scatterer is a perfectly conducting sphere. The exact Green's function and the electric far-field patterns of the scattering problem are constructed. For a small sphere, a closed-form approximation of the scattered wave field at the source of the incident spherical wave is obtained. These near-field results lead to the solution of an inverse problem. We also treat the same inverse problem using far-field results via the leading order term in the low-frequency asymptotic expansion of the scattering cross-section.
Keywords: chiral media; spherical scattered; point sources; inverse problems;

## 1. INTRODUCTION

In a homogeneous isotropic chiral media the electromagnetic fields are composed of left – circularly polarized (LCP) and right – circularly polarized (RCP) components, which have different wave numbers and independent directions of propagation.

The LCP and RCP components are assumed to be spherical Beltrami fields, [15], [16],[17], in practice, such wave fields are more readily realized.

The present authors have studied electromagnetic waves in chiral media generated by a point source in the vicinity of the scatterer. In particular, in [1], reciprocity, optical and general scattering theorems for point-source excitation in chiral media has been proved. In this article, we study two kinds of inverse problem. One involves far-field measurements,[3], and the other involves near-field measurements. Specifically, in the first case we measure the scattering cross-section for five point-source locations, while in the second case we measure the scattered field at the location of the point source.

Correspondence to: *Athanasiadis C,* Department of Mathematics, University of Athens, Panepistemiopolis, GR 15784 Athens, Greece, or *Berketis N.*, PhD, MSc Applied Mathematics, University of Athens – Department of Mathematics, Principal 5$^{ου}$ General Lyceum, Tataoulou and Medea st. 16232 - Byron -Athens, Greece.
1.E-mail: cathan@math.uoa.gr
2.E-mail: nberketis@gmail.com

Near-field inverse scattering, in which the incident field is generated by a point source, has been studied in [4] for acoustic and [5] for electromagnetic waves in an achiral environment, while results for far-field data can be found in, [5], [7], [10] and [11].

In Section 2, considering Bohren decomposition into suitable Beltrami fields, we formulate the direct scattering problem of a spherical electromagnetic wave by a perfectly conducting obstacle. This problem is well posed the existence and uniqueness has been proved in [6].

In Section 3, after expanding the incident field in terms of spherical wave functions, we obtain the exact solution of the scattering problem as well as an expansion for the electric far-field pattern.

In Section 4, we consider either LCP or RCP incidence and we obtain an approximation of the scattering cross-section. For the far-field experiments, we measure the scattering cross-section for various point-source locations.

Finally in Section 5, using near-field experiments, in which the scattered field is measured at the source, we solve the corresponding inverse scattering problem.

## 2. STATEMENT OF THE PROBLEM

Our goal is to study the direct and inverse scattering problems when the incident electromagnetic field is a time harmonic point – generated wave in a chiral medium and the scatterer is a perfectly conducting sphere of radius $a$ centered at the origin. The exterior space $(r=|\mathbf{r}|>a)$ is an infinite homogeneous isotropic chiral medium with chirality measure β, electric permittivity ε and magnetic permeability μ.

We consider a time harmonic spherical electromagnetic wave due to a point source at $P_0$ with position vector $\mathbf{r}_0$ with respect to an origin O in the vicinity of the scatterer. In order to define spherical electromagnetic fields $\mathbf{E}_{\mathbf{r}_0}, \mathbf{H}_{\mathbf{r}_0}$, we make use of the Bohren decomposition into Beltrami fields $\mathbf{Q}_{L,\mathbf{r}_0}$ and $\mathbf{Q}_{R,\mathbf{r}_0}$, as follows

$$\mathbf{E}_{\mathbf{r}_0} = \mathbf{Q}_{L,\mathbf{r}_0} + \mathbf{Q}_{R,\mathbf{r}_0}, \quad \mathbf{H}_{\mathbf{r}_0} = \frac{1}{i\eta}\left(\mathbf{Q}_{L,\mathbf{r}_0} - \mathbf{Q}_{R,\mathbf{r}_0}\right), \tag{2.1}$$

where $\eta = \left(\dfrac{\mu}{\varepsilon}\right)^{1/2}$ is the intrinsic impedance of the chiral medium. The Beltrami fields satisfy the equations

$$\nabla \times \mathbf{Q}_{L,\mathbf{r}_0} = \gamma_L \mathbf{Q}_{L,\mathbf{r}_0}, \quad \nabla \times \mathbf{Q}_{R,\mathbf{r}_0} = -\gamma_R \mathbf{Q}_{L,\mathbf{r}_0}, \tag{2.2}$$

where $\gamma_L$ and $\gamma_R$ are wave numbers given by

$$\gamma_L = \frac{k}{1-k\beta}, \quad \gamma_R = \frac{k}{1+k\beta}, \tag{2.3}$$

with $\kappa = \omega(\varepsilon\mu)^{1/2}$, $\omega$ being the angular frequency. The in deices L and R denote the LCP and RCP fields respectively. The spherical incident Beltrami fields with suitable normalization have the form, [1]

$$\mathbf{Q}_{L,\mathbf{r}_0}^{inc}(\mathbf{r}|\hat{\mathbf{p}}_L) = \frac{1}{2}\left[\tilde{\mathbf{I}} + \frac{1}{\gamma_L^2}\nabla\nabla + \frac{1}{\gamma_L}\nabla\times\tilde{\mathbf{I}}\right]\left(\frac{h_0(\gamma_L|\mathbf{r}-\mathbf{r}_0|)}{h_0(\gamma_L r_0)}\right)\cdot\hat{\mathbf{p}}_L \quad (2.4)$$

$$= \frac{2\pi r_0 e^{-i\gamma_L r_0}}{\gamma_L}\left[\gamma_L \tilde{\mathbf{G}}_{fs}(\mathbf{r},\mathbf{r}_0) + \nabla\times\tilde{\mathbf{G}}_{fs}(\mathbf{r},\mathbf{r}_0)\right]\cdot\hat{\mathbf{p}}_L,$$

$$\mathbf{Q}_{R,\mathbf{r}_0}^{inc}(\mathbf{r}|\hat{\mathbf{p}}_R) = \frac{1}{2}\left[\tilde{\mathbf{I}} + \frac{1}{\gamma_R^2}\nabla\nabla - \frac{1}{\gamma_R}\nabla\times\tilde{\mathbf{I}}\right]\left(\frac{h_0(\gamma_R|\mathbf{r}-\mathbf{r}_0|)}{h_0(\gamma_R r_0)}\right)\cdot\hat{\mathbf{p}}_R \quad (2.5)$$

$$= \frac{2\pi r_0 e^{-i\gamma_R r_0}}{\gamma_R}\left[\gamma_R \tilde{\mathbf{G}}_{fs}(\mathbf{r},\mathbf{r}_0) - \nabla\times\tilde{\mathbf{G}}_{fs}(\mathbf{r},\mathbf{r}_0)\right]\cdot\hat{\mathbf{p}}_R,$$

where $h_0(x) = h_0^1(x) = e^{ix}/(ix)$ is the zeroth-order spherical Hankel function of first kind, $\tilde{\mathbf{I}} = \hat{\mathbf{x}}\hat{\mathbf{x}} + \hat{\mathbf{\psi}}\hat{\mathbf{\psi}} + \hat{\mathbf{z}}\hat{\mathbf{z}}$ is the identity dyadic and $r_0 = |\mathbf{r}_0|$ and $\tilde{\mathbf{G}}_{fs}(\mathbf{r},\mathbf{r}_0)$ is the free space dyadic Green function , p.42 , [15] The constant unit vectors $\hat{\mathbf{p}}_L$ and $\hat{\mathbf{p}}_R$ satisfy the relations

$$\hat{\mathbf{r}}_0 \cdot \hat{\mathbf{p}}_L = \hat{\mathbf{r}}_0 \cdot \hat{\mathbf{p}}_R = 0, \quad \hat{\mathbf{r}}_0 \times \hat{\mathbf{p}}_L = i\hat{\mathbf{p}}_L, \quad , \quad \hat{\mathbf{r}}_0 \times \hat{\mathbf{p}}_R = -i\hat{\mathbf{p}}_R. \quad (2.6)$$

We note that when $r_0 \to \infty$, the incident electric field

$$\mathbf{E}_{\mathbf{r}_0}^{inc}(\mathbf{r}|\hat{\mathbf{p}}_L,\hat{\mathbf{p}}_R) = \mathbf{Q}_{L,\mathbf{r}_0}^{inc}(\mathbf{r}|\hat{\mathbf{p}}_L) + \mathbf{Q}_{R,\mathbf{r}_0}^{inc}(\mathbf{r}|\hat{\mathbf{p}}_R), \quad (2.7)$$

reduces to plane electric wave with direction of propagation $-\hat{\mathbf{r}}_0$ and polarizations $\hat{\mathbf{p}}_L, \hat{\mathbf{p}}_R$, since

$$\lim_{r_0\to\infty}\mathbf{Q}_{L,\mathbf{r}_0}^{inc}(\mathbf{r}|\hat{\mathbf{p}}_L) = \hat{\mathbf{p}}_L e^{-i\gamma_L \hat{\mathbf{r}}_0\cdot\mathbf{r}} = \mathbf{Q}_L^{inc}(\mathbf{r};-\hat{\mathbf{r}}_0,\hat{\mathbf{p}}_L), \quad (2.8)$$

$$\lim_{r_0\to\infty}\mathbf{Q}_{R,\mathbf{r}_0}^{inc}(\mathbf{r}|\hat{\mathbf{p}}_R) = \hat{\mathbf{p}}_R e^{-i\gamma_R \hat{\mathbf{r}}_0\cdot\mathbf{r}} = \mathbf{Q}_R^{inc}(\mathbf{r};-\hat{\mathbf{r}}_0,\hat{\mathbf{p}}_R), \quad (2.9)$$

We consider $\mathbf{E}_{\mathbf{r}_0}^{inc}$ is incident upon a perfectly conducting sphere of radius a. Then, we want to calculate the scattered electric field $\mathbf{E}_{\mathbf{r}_0}^{sc}$ which is the unique solution, [6], of the following exterior boundary value problem

$$\nabla\times\nabla\times\mathbf{E}_{\mathbf{r}_0}^{sc}(\mathbf{r}) - 2\gamma^2\beta\nabla\times\mathbf{E}_{\mathbf{r}_0}^{sc}(\mathbf{r}) - \gamma^2\mathbf{E}_{\mathbf{r}_0}^{sc}(\mathbf{r}) = \mathbf{0}, \; r > a \quad (2.10)$$

$$\hat{\mathbf{n}}\times\mathbf{E}_{\mathbf{r}_0}^{sc}(\mathbf{r}) = -\hat{\mathbf{n}}\times\mathbf{E}_{\mathbf{r}_0}^{inc}(\mathbf{r}), \; r = a \quad (2.11)$$

$$\hat{\mathbf{r}}\times\nabla\times\mathbf{E}_{\mathbf{r}_0}^{sc}(\mathbf{r}) - \beta\gamma^2\hat{\mathbf{r}}\times\mathbf{E}_{\mathbf{r}_0}^{sc}(\mathbf{r}) + \frac{i\gamma^2}{k}\mathbf{E}_{\mathbf{r}_0}^{sc}(\mathbf{r}) = o\left(\frac{1}{r}\right), \quad r\to\infty. \quad (2.12)$$

The radiation condition (2.12) is valued uniformly in all directions $\hat{\mathbf{r}}\in S^2$, where $S^2$ is the unit sphere in $\mathbb{R}^3$, $\hat{\mathbf{n}}$ is the outward normal unit vector on the scatterer and $\gamma^2 = \gamma_L\gamma_R$. The scattered electric field will be depended on the polarizations $\hat{\mathbf{p}}_L$ and $\hat{\mathbf{p}}_R$ and will have the decomposition

$$\mathbf{E}_{\mathbf{r}_0}^{sc}(\mathbf{r}|\hat{\mathbf{p}}_L,\hat{\mathbf{p}}_R) = \mathbf{Q}_{L,\mathbf{r}_0}^{sc}(\mathbf{r}|\hat{\mathbf{p}}_L,\hat{\mathbf{p}}_R) + \mathbf{Q}_{R,\mathbf{r}_0}^{sc}(\mathbf{r}|\hat{\mathbf{p}}_L,\hat{\mathbf{p}}_R), \quad (2.13)$$

where $\mathbf{Q}_{L,\mathbf{r}_0}^{sc}(\mathbf{r}|\hat{\mathbf{p}}_L,\hat{\mathbf{p}}_R)$ and $\mathbf{Q}_{R,\mathbf{r}_0}^{sc}(\mathbf{r}|\hat{\mathbf{p}}_L,\hat{\mathbf{p}}_R)$ are the corresponding scattered Beltrami fields which have the following behavior, when $r\to\infty$

$$\mathbf{Q}_{L,\mathbf{r}_0}^{sc}(\mathbf{r}|\hat{\mathbf{p}}_L,\hat{\mathbf{p}}_R) = h_0(\gamma_L r)\mathbf{g}_{L,\mathbf{r}_0}(\hat{\mathbf{r}}|\hat{\mathbf{p}}_L,\hat{\mathbf{p}}_R) + O\left(\frac{1}{r^2}\right), \quad r\to\infty, \quad (2.14)$$

$$\mathbf{Q}_{R,\mathbf{r}_0}^{sc}\left(\mathbf{r}|\hat{\mathbf{p}}_L,\hat{\mathbf{p}}_R\right) = h_0\left(\gamma_R r\right)\mathbf{g}_{R,\mathbf{r}_0}\left(\hat{\mathbf{r}}|\hat{\mathbf{p}}_L,\hat{\mathbf{p}}_R\right) + O\left(\frac{1}{r^2}\right), \quad r \to \infty, \qquad (2.15)$$

The functions $\mathbf{g}_{L,\mathbf{r}_0}$ and $\mathbf{g}_{R,\mathbf{r}_0}$ are the LCP and RCP far – field patterns respectively, [1]

If either a LCP or a RCP spherical electric wave $\mathbf{E}_{\mathbf{r}_0}^{inc}\left(\mathbf{r}|\hat{\mathbf{p}}_A\right)$, $A = L, R$, is incident upon the scatterer, then the scattering cross – section, is given by [3],

$$\sigma_{A,\mathbf{r}_0}^{sc} = \int_{S^2}\left[\frac{1}{\gamma_L^2}\left|\mathbf{g}_{L,\mathbf{r}_0}\left(\hat{\mathbf{r}}|\hat{\mathbf{p}}_A\right)\right|^2 + \frac{1}{\gamma_R^2}\left|\mathbf{g}_{R,\mathbf{r}_0}\left(\hat{\mathbf{r}}|\hat{\mathbf{p}}_A\right)\right|^2\right]ds(\hat{\mathbf{r}}). \qquad (2.16)$$

### 3. EXACT GREEN'S FUNCTION

We take spherical polar coordinates $(r,\theta,\varphi)$ where $\theta \in [0,\pi]$ and $\varphi \in [0,2\pi]$, with the origin at the centre of the spherical scatterer, so that the point source is at $r = r_0$, $\theta = 0$. Thus, $\mathbf{r}_0 = r_0\hat{\mathbf{z}}$, $\hat{\mathbf{p}}_L = \frac{1}{\sqrt{2}}(\hat{\mathbf{x}} - i\hat{\boldsymbol{\psi}})$ and $\hat{\mathbf{p}}_R = \frac{1}{\sqrt{2}}(\hat{\mathbf{x}} + i\hat{\boldsymbol{\psi}})$, where $\hat{\mathbf{x}}, \hat{\boldsymbol{\psi}}$ and $\hat{\mathbf{z}}$ are unit vectors in the $x, \psi$ and $z$ directions, respectively. Using spherical vector wave functions (see 13.3.68 – 13.3.70 of [19], [21] and p.49-52 of [15] ) and taking into account (2.4), (2.5), we obtain

$$\mathbf{Q}_{L,\mathbf{r}_0}^{inc}\left(\mathbf{r}|\hat{\mathbf{p}}_L\right) = \sum_{n=1}^{\infty} B_n^L \left\{\mathbf{L}_{o1n}^{(1)}(\gamma_L\mathbf{r}) + i\mathbf{L}_{e1n}^{(1)}(\gamma_L\mathbf{r})\right\}, \qquad (3.1)$$

or

$$\mathbf{Q}_{R,\mathbf{r}_0}^{inc}\left(\mathbf{r}|\hat{\mathbf{p}}_R\right) = \sum_{n=1}^{\infty} B_n^R \left\{\mathbf{R}_{o1n}^{(1)}(\gamma_R\mathbf{r}) - i\mathbf{R}_{e1n}^{(1)}(\gamma_R\mathbf{r})\right\}, \qquad (3.2)$$

where

$$B_n^A = \frac{1}{2\sqrt{2}h_0\left(\gamma_A r_0\right)}\frac{2n+1}{n(n+1)}H_n\left(\gamma_A r_0\right) \text{ and } H_n\left(\gamma_A r_0\right) = h_n\left(\gamma_A r_0\right) - i\tilde{h}_n\left(\gamma_A r_0\right), \quad (3.3)$$

with $A = L,R$, for $r < r_0$. The $h_n$ is a spherical Hankel function of first order, $\tilde{h}(x) = x^{-1}h_n(x) + h_n'(x)$, and $\mathbf{L}_{s1n}^{(\rho)}$ and $\mathbf{R}_{s1n}^{(\rho)}$, with $s = o,e$, are the spherical functions, [15],

$$\mathbf{L}_{s1n}^{(\rho)}(\gamma_L\mathbf{r}) = \mathbf{M}_{s1n}^{(\rho)}(\gamma_L\mathbf{r}) + \mathbf{N}_{s1n}^{(\rho)}(\gamma_L\mathbf{r}), \quad \mathbf{R}_{s1n}^{(\rho)}(\gamma_R\mathbf{r}) = \mathbf{M}_{s1n}^{(\rho)}(\gamma_R\mathbf{r}) - \mathbf{N}_{s1n}^{(\rho)}(\gamma_R\mathbf{r}), \qquad (3.4)$$

where $\rho = 1,3$, the $\mathbf{M}_{s1n}^{(\rho)}$ and $\mathbf{N}_{s1n}^{(\rho)}$ are known spherical vector function,[19]. The scattered electric field that comes from a LCP incident field or a RCP incident field has a similar expansion to (3.1) or to (3.2)

$$\mathbf{E}^{sc}_{\mathbf{r}_0}\left(\mathbf{r}|\hat{\mathbf{p}}_L\right) = \mathbf{Q}^{sc}_{L,\mathbf{r}_0}\left(\mathbf{r}|\hat{\mathbf{p}}_L\right) + \mathbf{Q}^{sc}_{R,\mathbf{r}_0}\left(\mathbf{r}|\hat{\mathbf{p}}_L\right)$$

$$= \sum_{n=1}^{\infty} B_n^L a_n^L \left\{ \mathbf{L}^{(3)}_{o1n}(\gamma_L \mathbf{r}) + i\mathbf{L}^{(3)}_{e1n}(\gamma_L \mathbf{r}) \right\} \qquad (3.5)$$

$$+ \sum_{n=1}^{\infty} B_n^L a_n^R \left\{ \mathbf{R}^{(3)}_{o1n}(\gamma_R \mathbf{r}) + i\mathbf{R}^{(3)}_{e1n}(\gamma_R \mathbf{r}) \right\},$$

or

$$\mathbf{E}^{sc}_{\mathbf{r}_0}\left(\mathbf{r}|\hat{\mathbf{p}}_R\right) = \mathbf{Q}^{sc}_{L,\mathbf{r}_0}\left(\mathbf{r}|\hat{\mathbf{p}}_R\right) + \mathbf{Q}^{sc}_{R,\mathbf{r}_0}\left(\mathbf{r}|\hat{\mathbf{p}}_R\right)$$

$$= \sum_{n=1}^{\infty} B_n^R b_n^L \left\{ \mathbf{L}^{(3)}_{o}(\gamma_L \mathbf{r}) - i\mathbf{L}^{(3)}_{e}(\gamma_L \mathbf{r}) \right\} \qquad (3.6)$$

$$+ \sum_{n=1}^{\infty} B_n^R b_n^R \left\{ \mathbf{R}^{(3)}_{o}(\gamma_R \mathbf{r}) - i\mathbf{R}^{(3)}_{e}(\gamma_R \mathbf{r}) \right\}.$$

Using the boundary condition (2.11) on $r = a$, we obtain

$$a_n^L = -\frac{j_n(\gamma_L a)\tilde{h}_n(\gamma_R a) + \tilde{j}_n(\gamma_L a)h_n(\gamma_R a)}{h_n(\gamma_L a)\tilde{h}_n(\gamma_R a) + \tilde{h}_n(\gamma_L a)h_n(\gamma_R a)}, \qquad (3.7)$$

and

$$a_n^R = -\frac{j_n(\gamma_L a)\tilde{h}_n(\gamma_L a) - \tilde{j}_n(\gamma_L a)h_n(\gamma_L a)}{h_n(\gamma_L a)\tilde{h}_n(\gamma_R a) + \tilde{h}_n(\gamma_L a)h_n(\gamma_R a)}, \qquad (3.8)$$

or

$$b_n^L = -\frac{j_n(\gamma_R a)\tilde{h}_n(\gamma_R a) - \tilde{j}_n(\gamma_R a)h_n(\gamma_R a)}{h_n(\gamma_L a)\tilde{h}_n(\gamma_R a) + \tilde{h}_n(\gamma_L a)h_n(\gamma_R a)}, \qquad (3.9)$$

and

$$b_n^R = -\frac{j_n(\gamma_R a)\tilde{h}_n(\gamma_L a) + \tilde{j}_n(\gamma_R a)h_n(\gamma_L a)}{h_n(\gamma_L a)\tilde{h}_n(\gamma_R a) + \tilde{h}_n(\gamma_L a)h_n(\gamma_R a)}. \qquad (3.10)$$

Using the asymptotic forms, [15],

$$\mathbf{L}^{(3)}_{s1n}(\gamma_L \mathbf{r}) \sim \sqrt{n(n+1)}(-i)^n \mathbf{f}^L_{s1n}(\hat{\mathbf{r}}) h_0(\gamma_L r), \qquad (3.11)$$

$$\mathbf{R}^{(3)}_{s1n}(\gamma_R \mathbf{r}) \sim \sqrt{n(n+1)}(-i)^n \mathbf{f}^R_{s1n}(\hat{\mathbf{r}}) h_0(\gamma_R r), \qquad (3.12)$$

where let us introduce LCP Beltrami angular $\mathbf{f}^L_{s1n}(\hat{\mathbf{r}})$, and RCP Beltrami angular $\mathbf{f}^R_{s1n}(\hat{\mathbf{r}})$, [15], satisfy by relations

$$\mathbf{f}^L_{s1n}(\hat{\mathbf{r}}) = \mathbf{C}_{s1n}(\hat{\mathbf{r}}) + i\mathbf{B}_{s1n}(\hat{\mathbf{r}}), \quad \mathbf{f}^R_{s1n}(\hat{\mathbf{r}}) = \mathbf{C}_{s1n}(\hat{\mathbf{r}}) - i\mathbf{B}_{s1n}(\hat{\mathbf{r}}), \qquad (3.13)$$

we calculate the electric far-field patterns:

$$\mathbf{g}^{sc}_{A,\mathbf{r}_0}(\hat{\mathbf{r}}|\hat{\mathbf{p}}_L) = \sum_{n=1}^{\infty} \frac{(2n+1)(-i)^{n-1}}{2\sqrt{2n(n+1)}} \frac{H_n(\gamma_L r_0)}{h_0(\gamma_L r_0)} a_n^A \left\{ \mathbf{f}^A_{e1n}(\hat{\mathbf{r}}) - i\mathbf{f}^A_{o1n}(\hat{\mathbf{r}}) \right\}, \qquad (3.14)$$

or

$$\mathbf{g}_{A,\mathbf{r}_0}^{sc}(\hat{\mathbf{r}}|\hat{\mathbf{p}}_R) = \sum_{n=1}^{\infty} \frac{(2n+1)(-i)^{n-1}}{2\sqrt{2n(n+1)}} \frac{H_n(\gamma_R r_0)}{h_0(\gamma_R r_0)} b_n^A \{-\mathbf{f}_{e1n}^A(\hat{\mathbf{r}}) - i\mathbf{f}_{o1n}^A(\hat{\mathbf{r}})\}, \quad (3.15)$$

with $A = L, R$.

## 4. A FAR-FIELD INVERSE PROBLEM

So far, all of our formulas are exact. In the asymptotic results to follow, there are three parameters $\gamma_A a$, with $A = L, R$ and $\tau = a/r_0$. We note that the geometrical parameter $\tau$ must satisfy $0 < \tau < 1$ because the point source is outside of the sphere.

We assume that $|\gamma_A a| \ll 1$, as well; that is we make the so-called low-frequency assumption. From (3.7), (3.8), (3.10), (3.11), we obtain

$$a_n^L \sim \frac{1+\beta k}{2i\zeta_n^2(2n+1)}(\gamma_L a)^{2n+1}, \quad a_n^R \sim -\frac{i}{2n\zeta_n^2} \frac{(1-\beta k)^{n+2}}{(1+\beta k)^{n+1}}(\gamma_L a)^{2n+1}, \gamma_L a \to 0, \quad (4.1)$$

or

$$b_n^L \sim -\frac{i}{2n\zeta_n^2} \frac{(1+\beta k)^{n+2}}{(1-\beta k)^{n+1}}(\gamma_R a)^{2n+1}, \gamma_R a \to 0, \quad b_n^R \sim -\frac{i(1-\beta k)}{2\zeta_n^2(2n+1)}(\gamma_R a)^{2n+1}, \quad (4.2)$$

where $\zeta_n = 1 \cdot 3 \cdot 5 \cdots (2n-1) = (2n)!/(2^n n!)$. In particular,

$$\begin{cases} a_1^L = -i\frac{1+\beta k}{6}(\gamma_L a)^3 + O((\gamma_L a)^5) \\ a_2^L = -i\frac{1+\beta k}{90}(\gamma_L a)^5 + O((\gamma_L a)^7) \end{cases}, \gamma_L a \to 0, \quad (4.3)$$

$$\begin{cases} a_1^R = -\frac{i(1-\beta k)^3}{2(1+\beta k)^2}(\gamma_L a)^3 + O((\gamma_L a)^5) \\ a_2^R = -\frac{i(1-\beta k)^4}{36(1+\beta k)^3}(\gamma_L a)^5 + O((\gamma_L a)^7) \end{cases}, \gamma_L a \to 0, \quad (4.4)$$

or

$$\begin{cases} b_1^L = -\frac{i(1+\beta k)^3}{2(1-\beta k)^2}(\gamma_R a)^3 + O((\gamma_R a)^5) \\ b_2^L = -\frac{i(1+\beta k)^4}{36(1-\beta k)^3}(\gamma_R a)^5 + O((\gamma_R a)^7) \end{cases}, \gamma_R a \to 0, \quad (4.5)$$

$$\begin{cases} b_1^R = -\dfrac{i(1-\beta k)}{6}(\gamma_R a)^3 + O\left((\gamma_R a)^5\right) \\ b_2^R = -\dfrac{i(1-\beta k)}{90}(\gamma_R a)^5 + \left((\gamma_R a)^7\right) \end{cases}, \gamma_R a \to 0. \qquad (4.6)$$

as $\gamma_A a \to 0$. In order to calculate $\mathbf{g}_{A,\mathbf{r}_0}^{sc}$, $A = L, R$ with an error of $O\left((\gamma_A a)^4\right)$ we only need the following, [19]

$$\mathbf{C}_{o11}(\hat{\mathbf{r}}) = \frac{1}{\sqrt{2}}\left[\hat{\boldsymbol{\theta}}\cos\varphi - \hat{\boldsymbol{\varphi}}\cos\theta\sin\varphi\right], \mathbf{B}_{o11}(\hat{\mathbf{r}}) = -\frac{1}{\sqrt{2}}\left[\hat{\boldsymbol{\theta}}\cos\theta\sin\varphi + \hat{\boldsymbol{\varphi}}\cos\varphi\right], \quad (4.7)$$

$$\mathbf{B}_{e11}(\hat{\mathbf{r}}) = \frac{1}{\sqrt{2}}\left[\hat{\boldsymbol{\theta}}\cos\theta\sin\varphi - \hat{\boldsymbol{\varphi}}\sin\varphi\right], \mathbf{C}_{e11}(\hat{\mathbf{r}}) = \frac{1}{\sqrt{2}}\left[-\hat{\boldsymbol{\theta}}\sin\varphi - \hat{\boldsymbol{\varphi}}\cos\theta\cos\varphi\right], \quad (4.8)$$

$$\mathbf{C}_{o12}(\hat{\mathbf{r}}) = (3/2)^{1/2}\left\{\hat{\boldsymbol{\theta}}\cos\theta\cos\varphi - \hat{\boldsymbol{\varphi}}\cos 2\theta\sin\varphi\right\}, \qquad (4.9)$$

$$\mathbf{B}_{e12}(\hat{\mathbf{r}}) = (3/2)^{1/2}\left\{\hat{\boldsymbol{\theta}}\cos 2\theta\cos\varphi - \hat{\boldsymbol{\varphi}}\cos\theta\sin\varphi\right\}, \qquad (4.10)$$

$$\mathbf{B}_{o12}(\hat{\mathbf{r}}) = (3/2)^{1/2}\left\{\hat{\boldsymbol{\varphi}}\cos\theta\cos\varphi + \hat{\boldsymbol{\theta}}\cos 2\theta\sin\varphi\right\}, \qquad (4.11)$$

$$\hat{\mathbf{C}}_{e12}(\hat{\mathbf{r}}) = (3/2)^{1/2}\left\{-\hat{\boldsymbol{\varphi}}\cos 2\theta\cos\varphi - \hat{\boldsymbol{\theta}}\cos\theta\sin\varphi\right\}, \qquad (4.12)$$

So from (3.14) and (3.15), we finally obtain

$$\mathbf{g}_{A,\mathbf{r}_0}^{sc}\left(\hat{\mathbf{r}}\big|\hat{\mathbf{p}}_A\right) = \varpi_A \frac{(1-\varpi_A \beta k)}{8}\left\{-(\gamma_A a)\tau^2\left\{\mathbf{f}_{e11}^A(\hat{\mathbf{r}}) + \varpi_A i\mathbf{f}_{o11}^A(\hat{\mathbf{r}})\right\}\right.$$
$$+(\gamma_A a)^2\left\{2i\tau\left\{\mathbf{f}_{e11}^A(\hat{\mathbf{r}}) + \varpi_A i\mathbf{f}_{o11}^A(\hat{\mathbf{r}})\right\} + \frac{i}{\sqrt{3}}\tau^3\left\{\mathbf{f}_{e12}^A(\hat{\mathbf{r}}) + \varpi_A i\mathbf{f}_{o12}^A(\hat{\mathbf{r}})\right\}\right\} \qquad (4.13)$$
$$\left.+(\gamma_A a)^3\left\{2\left\{\mathbf{f}_{e11}^A(\hat{\mathbf{r}}) + \varpi_A i\mathbf{f}_{o11}^A(\hat{\mathbf{r}})\right\} + \frac{4}{3\sqrt{3}}\tau^2\left\{\mathbf{f}_{e12}^A(\hat{\mathbf{r}}) + \varpi_A i\mathbf{f}_{o12}^A(\hat{\mathbf{r}})\right\}\right\}\right\} + O\left((\gamma_A a)^4\right),$$

and

$$\mathbf{g}_{A,\mathbf{r}_0}^{sc}\left(\hat{\mathbf{r}}\big|\hat{\mathbf{p}}_{A^c}\right) = \varpi_A \frac{3(1-\varpi_A \beta k)^2}{8(1+\varpi_A \beta k)}(\gamma_A a)\tau^2\left\{\mathbf{f}_{e11}^A(\hat{\mathbf{r}}) - \varpi_A i\mathbf{f}_{o11}^A(\hat{\mathbf{r}})\right\}$$
$$+(\gamma_A a)^2\left\{-\varpi_A \frac{3i(1-\varpi_A \beta k)}{4}\tau\left\{\mathbf{f}_{e11}^R(\hat{\mathbf{r}}) - \varpi_A i\mathbf{f}_{o11}^R(\hat{\mathbf{r}})\right\}\right.$$
$$\left.-\varpi_A \frac{5i}{16\sqrt{3}}\frac{(1-\varpi_A \beta k)^2}{1+\varpi_A \beta k}\tau^3\left\{\mathbf{f}_{e12}^R(\hat{\mathbf{r}}) - \varpi_A i\mathbf{f}_{o12}^R(\hat{\mathbf{r}})\right\}\right\} \qquad (4.14)$$
$$+(\gamma_A a)^3\left\{-\varpi_A \frac{3(1+\varpi_A \beta k)}{4}\left\{\mathbf{f}_{e11}^R(\hat{\mathbf{r}}) - \varpi_A i\mathbf{f}_{o11}^R(\hat{\mathbf{r}})\right\}\right.$$
$$\left.-\varpi_A \frac{5(1-\varpi_A \beta k)}{12\sqrt{3}}\tau^2\left\{\mathbf{f}_{e12}^R(\hat{\mathbf{r}}) - \varpi_A i\mathbf{f}_{o12}^R(\hat{\mathbf{r}})\right\}\right\} + O\left((\gamma_R a)^4\right),$$

where $\varpi_A = \begin{cases} -1, & A = L \\ 1, & A = R \end{cases}$ and if $A = L, R$ at that case $A^c = R, L$.

Now for the scattering cross – section, by LCP or RCP spherical Beltrami fields, that is given by the forma (2.16), given by the relations

$$\sigma_{L,r_0}^{sc} = \frac{(1+\beta k)^2}{64}(\pi a^2)\left\{\frac{16}{3}\tau^4 + (\gamma_L a)^2\left(\frac{64}{3}\tau^2 + 16\tau^6\right)\right.$$
$$\left.+(\gamma_L a)^4\left(\frac{64}{3} + \frac{16}{9}\tau^4\right)\right\} + O\left((\gamma_L a)^6\right) + \pi a^2\left\{\frac{3(1-\beta k)^4}{4(1+\beta k)^2}\tau^4\right.$$
$$+(\gamma_R a)^2\left(3(1-\beta k)^2\tau^2 + \frac{5}{16}\frac{(1-\beta k)^4}{(1+\beta k)^2}\tau^6\right)$$
$$\left.+(\gamma_R a)^4\left(3(1+\beta k)^2 + \frac{5(1-\beta k)}{9}\tau^4\right)\right\} + O\left((\gamma_R a)^6\right), \quad (4.15)$$

or

$$\sigma_{R,r_0}^{sc} = \frac{(1-\beta k)^2}{64}(\pi a^2)\left\{\frac{16}{3}\tau^4 + (\gamma_R a)^2\left(\frac{64}{3}\tau^2 + 16\tau^6\right)\right.$$
$$\left.+(\gamma_R a)^4\left(\frac{64}{3} + \frac{16}{9}\tau^4\right)\right\} + O\left((\gamma_R a)^6\right)$$
$$+\pi a^2\left\{\frac{3(1+\beta k)^4}{4(1-\beta k)^2}\tau^4 + (\gamma_L a)^2\left(3(1+\beta k)^2\tau^2 + \frac{5}{16}\frac{(1+\beta k)^4}{(1-\beta k)^2}\tau^6\right)\right.$$
$$\left.+(\gamma_L a)^4\left(3(1-\beta k)^2 + \frac{5(1+\beta k)}{9}\tau^4\right)\right\} + O\left((\gamma_L a)^6\right). \quad (4.16)$$

In the special case $r_0 \to \infty$ ($\tau \to 0$), by the relations (4.15), (4.16) we obtain

$$\sigma = (\pi a^2)\left\{\frac{(1+\beta k)^2}{3} + \frac{3(1-\beta k)^4}{(1+\beta k)^2}\right\}(\gamma_L a)^4 + O\left((\gamma_L a)^6\right), \gamma_L a \to 0, \quad (4.17)$$

or

$$\sigma = (\pi a^2)\left\{\frac{(1-\beta k)^2}{3} + \frac{3(1+\beta k)^4}{(1-\beta k)^2}\right\}(\gamma_R a)^4 + O\left((\gamma_R a)^6\right), \gamma_R a \to 0, \quad (4.18)$$

likewise in the case $\gamma_L a \to 0$ and $\gamma_R a \to 0$, by the relations (4.15), (4.16) we obtain

$$\sigma_{A,r_0}^{sc} = \frac{1}{4}f_A(\beta,k)(\pi a^2)(a/r_0)^4, \quad (4.19)$$

where $A = L, R$, with

$$f_A(\beta,k) = \frac{(1-\varpi_A\beta k)^2}{3} + \frac{3(1+\varpi_A\beta k)^4}{(1-\varpi_A\beta k)^2}, \quad A = L, R. \tag{4.20}$$

Choose a Cartesian coordinate system $Oxyz$, and five point-source locations, namely $(0,0,0), (l,0,0), (0,l,0), (0,0,l)$ and $(0,0,2l)$, which are at (unknown) distances $r_0, r_1, r_2, r_3$ and $r_4$, respectively, from the sphere's centre. The parameter $l$ is a chosen fixed length. For each location, measure the leading-order term in the low-frequency expansion of the scattering cross-section. Thus, our five measurements are

$$m_j = \frac{1}{4} f_A(\beta,k) \pi a^2 \left(\frac{a}{r_j}\right)^4, \quad j = 0,1,2,3,4, \tag{4.21}$$

Dimensionless quantities related to $m_j$ are

$$\gamma_j = \frac{l}{\sqrt{m_j}} = 2\sqrt{\frac{1}{f_A(\beta,k)\pi}} \frac{l}{a} \left(\frac{r_j}{a}\right)^2, \quad \text{where} \quad j = 0,1,2,3,4, \tag{4.22}$$

equivalently, we obtain

$$r_j^2 = \frac{1}{2}\sqrt{f_A(\beta,k)\pi} \frac{a^3}{l} \gamma_j, \text{ where } j = 0,1,2,3,4, \tag{4.23}$$

There are six unknowns namely $r_0, r_1, r_2, r_3, r_4$ and $a$. Furthermore, $r_0, r_3$ and $r_4$ are related using the cosine rule, $r_4^2 + r_0^2 = 2r_3^2 + 2l^2$. So, we can find the six unknowns. The centre of the spherical scatterer is obtained from the intersection of the four spheres centers at $(0,0,0), (l,0,0), (0,l,0)$ and $(0,0,l)$, with corresponding radius $r_0, r_1, r_2, r_3$ respectively.

## 5. A NEAR – FIELD INVERSE PROBLEMS

The scattered field at the source point is given by setting $\mathbf{r} = \mathbf{r}_0$ in (3.5) or (3.6) and using the relations, [19], (p. 1887), we obtain

$$\mathbf{E}_{\mathbf{r}_0}^{sc}(\mathbf{r}_0|\hat{\mathbf{p}}_L) = \frac{1}{4\sqrt{2}} \Bigg\{ -\frac{1+\beta k}{2}\left(\sum_{n=1}^{\infty}\tau^{2n+1} + \frac{1}{(\gamma_L r_0)^2}\sum_{n=1}^{\infty}n^2\tau^{2n+1}\right)$$
$$+ \frac{(1-\beta k)}{2}\left(-\sum_{n=1}^{\infty}\frac{(2n+1)}{n}\tau^{2n+1} - \frac{i}{\gamma_L r_0}\sum_{n=1}^{\infty}(2n+1)\tau^{2n+1} - \frac{i}{\gamma_R r_0}\sum_{n=1}^{\infty}(2n+1)\tau^{2n+1} \right. \tag{5.1}$$
$$\left. + \frac{1}{(\gamma_L r_0)(\gamma_R r_0)}\sum_{n=1}^{\infty}n(2n+1)\tau^{2n+1}\right)\Bigg\} \cdot \hat{\mathbf{p}}_L,$$

as $\gamma_L a \to 0$ and $\gamma_R a \to 0$ for fixed $\tau = a/r_0$. Summing the infinite series (recall that $0 \le \tau < 1$), we obtain

$$\mathbf{E}_{\mathbf{r}_0}^{sc}(\mathbf{r}_0|\hat{\mathbf{p}}_L) \sim \frac{(1+\beta k)}{2(\gamma_L a)^2} \frac{\tau^5}{(1-\tau^2)^3} \cdot \hat{\mathbf{p}}_L, \tag{5.2}$$

Finally, for $\gamma_A a \to 0$ we can obtain

$$\left|\mathbf{E}_{A,\mathbf{r}_0}^{sc}\left(\mathbf{r}_0|\hat{\mathbf{p}}_A\right)\right| \sim \frac{(1-\varpi_A\beta k)}{2(\gamma_A a)^2}\frac{\tau^5}{(1-\tau^2)^3}, \gamma_A a \to 0, A = L, R, \qquad (5.3)$$

This gives the electric of the scattered field at the source, for a sphere that is small compared to a wavelength. Let us now formulate a simple inverse problem. Thus, we consider measurements of the scattered field at the same five source points as in Section 4. We set

$$M_j = \frac{2(\gamma_L\gamma_R)l}{(1+\varpi_A\beta k)}\left|\mathbf{E}_{A,\mathbf{r}_j}^{sc}\left(\mathbf{r}_j|\hat{\mathbf{p}}_A\right)\right|, \qquad (5.4)$$

$$\rho_j = \frac{r_j}{l} \text{ και } b = \sqrt{\frac{a}{l}}, \qquad (5.5)$$

then we obtain

$$M_j = \frac{b^5\rho_j}{\left(\rho_j^2 - b^4\right)^3}, \qquad (5.6)$$

where $j = 0,1,2,3,4$. Thus, as before, we have five measurements with six $\left(\rho_0, \rho_1, \rho_2, \rho_3, \rho_4 \text{ and } b\right)$ unknowns, and the cosine-rule constraint, $\rho_4 = 2 + 2\rho_3 - \rho_0$. We also have $\rho_j > b^2 > 0$ and $r_j = \rho_j l$. We can write (5.6) as

$$\rho_j^6 - 3\rho_j^4 b^4 + 3\rho_j^2 b^8 - b^{12} - b^5\frac{\rho_j}{M_j} = 0, \qquad (5.7)$$

όπου $j = 0,1,2,3,4$, which is a cubic equation for $\rho_j$ if a is known. If a is not known, one has to solve the six algebraic equations, for the six unknowns. Analytical progress is possible if a is known to be small, so that $a \ll l$. Then, one can approximate (5.6) by

$$M_j = \frac{b^5}{\rho_j^5} \text{ with } j = 0,1,2,3,4. \qquad (5.8)$$

It follows that a can then be obtained from (5.8). Here, we have discussed a simple but genuinely near-field inverse problem. This is perhaps a more natural and realizable experiment. It is similar to, but more complicated than, the analogous acoustic problem analyzed in [4].

## 6. APPENDIX
### SPHERICAL VECTOR WAVE FUNCTIONS

In this appendix we include for convenience the definitions of the spherical vector wave functions used in the paper. Solutions to the vector wave (Helmholtz) equations

$$\nabla \times (\nabla \times \mathbf{E}(\mathbf{r})) - k^2 \mathbf{E}(\mathbf{r}) = 0, \tag{6.1}$$

are constructed using radial amplitudes and three vector spherical harmonics [11],[17],[19],[20]. The vector spherical harmonics may be expressed in terms of the scalar spherical harmonics,

$$\mathbf{P}_{s1n}(\theta, \varphi) = \hat{\mathbf{r}} Y_{s1n}(\theta, \varphi), \tag{6.2}$$

$$\mathbf{B}_{s1n}(\theta, \varphi) = r\left\{n(n+1)^{-1/2} \nabla Y_{s1n}(\theta, \varphi)\right\}, \tag{6.3}$$

$$\mathbf{C}_{s1n}(\theta, \varphi) = \left\{n(n+1)^{-1/2} \nabla \times \left[\mathbf{r} Y_{s1n}(\theta, \varphi)\right]\right\}, \tag{6.4}$$

where $s = e$ or $o$ (even or odd). Here, the spherical harmonics are defined by

$$Y_{e1n}(\theta, \varphi) = P_n^1(\cos\theta)\cos\varphi \text{ and } Y_{o1n}(\theta, \varphi) = P_n^1(\cos\theta)\sin\varphi, \tag{6.5}$$

with orthogonality relation

$$\int_0^\pi d\theta \sin\theta \int_0^{2\pi} d\varphi Y_n^m(\theta,\varphi) Y_{n'}^{-m'}(\theta,\varphi) = (-1)^m \frac{4\pi}{2n+1}\delta_{mm'}\delta_{nn'}, \tag{6.6}$$

where $P_n^1(\omega) = (1-\omega^2)^{1/2} P_n'(\omega)$, an associated Legendre is function and $P_n(\omega)$ is a Legendre polynomial. The first few orders of associated Legendre functions are as follows:
For $n = 0$ and $n = 1$, we have

$$P_0^0(\cos\theta) = 1, \tag{6.7}$$

$$P_1^{-1}(\cos\theta) = \frac{1}{2}\sin\theta, \ P_1^0(\cos\theta) = \cos\theta, \tag{6.8}$$

$$P_1^1(\cos\theta) = -\sin\theta, \ P_1^1(\cos\theta) = -\sin\theta. \tag{6.9}$$

So, for $n = 1, 2, \cdots$, we have the spherical vector wave functions of the first kind

$$\begin{aligned}\mathbf{M}_{s1n}^{(1)}(\mathbf{r}) &= curl\left[\mathbf{r} j_n(kr) Y_{s1n}(\theta,\varphi)\right] \\ &= \sqrt{n(n+1)}\mathbf{C}_{s1n}(\theta,\varphi) j_n(kr),\end{aligned} \tag{6.10}$$

$$\begin{aligned}\mathbf{N}_{s1n}^{(1)}(\mathbf{r}) &= \frac{1}{k} curl\mathbf{M}_{s1n}^{(1)} = \sqrt{n(n+1)}\mathbf{B}_{s1n}(\theta,\varphi)(kr)^{-1}(rj_n(kr))' \\ &+ n(n+1)\mathbf{P}_{s1n}(\theta,\varphi)(kr)^{-1} j_n(kr).\end{aligned} \tag{6.11}$$

The corresponding spherical vector wave function of the third, $\mathbf{M}_{s1n}^{(3)}$ and $\mathbf{N}_{s1n}^{(3)}$, are obtained by using the spherical Hankel functions $h_n(kr)$ instead of $j_n(kr)$ in the above definitions.